\documentclass[aps,prd,superscriptaddress,preprintnumbers]{revtex4}

\usepackage{array}
\usepackage{dcolumn}
\usepackage{graphicx}
\usepackage{latexsym}
\usepackage{hyperref}

\def\url#1{\mbox{\href{#1}{\sf #1}}}
\newcommand{\cfrac}[2]{\textstyle{\frac{#1}{#2}}}

\begin{document}

\preprint{NSF-ITP-03-08\hspace*{330pt}FERMILAB-Pub-03/018-T}

\title{Neutrino Observatories  Can Characterize \\ Cosmic Sources and Neutrino Properties}


\author{Gabriela Barenboim}
\email[E-mail: ]{gabriela@fnal.gov}
\affiliation{Theoretical Physics Department,  Fermi National 
Accelerator Laboratory, P.O.\ Box 500, Batavia, IL 60510}
\author{Chris Quigg}
\email[E-mail: ]{quigg@fnal.gov}
\affiliation{Theoretical Physics Department,  Fermi National 
Accelerator Laboratory, P.O.\ Box 500, Batavia, IL 60510}
\affiliation{Kavli Institute for Theoretical Physics, University of California, Santa Barbara, CA 93106}

\date{\today}

\begin{abstract}
Neutrino telescopes that measure relative fluxes of ultrahigh-energy
$\nu_{e}, \nu_{\mu}, \nu_{\tau}$ can give information about the
location and characteristics of sources, about neutrino mixing, and
can test for neutrino instability and for departures from \textsf{CPT}
invariance in the neutrino sector. We investigate consequences of neutrino mixing for the neutrino flux arriving at Earth, and consider how terrestrial measurements can characterize distant sources. We contrast mixtures that arise from neutrino oscillations with those signaling neutrino decays. We stress the importance of measuring $\nu_{e}, \nu_{\mu}, \nu_{\tau}$ fluxes in neutrino observatories.
\end{abstract}

\pacs{96.40.Tv Neutrinos and muons; 
cosmic rays \\14.60.Pq Neutrino mass and mixing \\ 13.35.Hb Decays of heavy 
neutrinos \\  11.30.Er \textsf{C, P, T} and other discrete symmetries}

\maketitle

\section{Introduction \label{intro}}
Neutrino telescopes promise to probe the deepest reaches of stars,
galaxies, and exotic structures in the
cosmos~\cite{MannLear,francis,francis03,GQRSapp}.  Unlike
charged particles, neutrinos arrive on a direct line from their
source, undeflected by magnetic fields.  Unlike photons, neutrinos
interact weakly, so they can penetrate thick columns of
matter~\footnote{The interaction length of a 1-TeV neutrino is about
2.5 million kilometers of water, or 250 kton/cm$^{2}$, whereas
high-energy photons are blocked by a few hundred g/cm$^{2}$.}. 
It is plausible that ultrahigh-energy extraterrestrial neutrinos will emerge from the atmospheric-neutrino background at energies between 1 and 10~TeV.
Prospecting for extraterrestrial neutrino sources leads the science
agenda for neutrino observatories, with characterizing the sources and
the processes that operate within them to follow. The fluxes of extraterrestrial neutrinos may, in addition, offer clues to the properties of neutrinos themselves.

The task of developing diagnostics for neutrino sources by measuring
relative fluxes of electron-, muon-, and tau-neutrinos is complicated
by neutrino oscillations.  Prominent among expected sources is the diffuse flux of neutrinos produced in the jets of active galactic nuclei such as the TeV-gamma sources Mkn~421 and 501, which are some 140~Mpc distant from Earth. The vacuum oscillation length, $L_{\mathrm{osc}} = 4\pi E_{\nu}/|\Delta m^2|$, is short compared with such intergalactic distances, so neutrinos oscillate many times between source and detector. For $|\Delta m^2| = 10^{-5}\hbox{ eV}^2$, for example, the oscillation length is $L_{\mathrm{osc}}\approx 2.5 \times 10^{-24}\hbox{ Mpc}\cdot (E_{\nu}/1\hbox{ eV})$, a fraction of a megaparsec even for $10^{20}$-eV neutrinos. The fluxes $\Phi = \{\varphi_{e},
\varphi_{\mu}, \varphi_{\tau}\}$ that arrive at Earth are not
identical to the source fluxes $\Phi^{0} = \{\varphi_{e}^{0},
\varphi_{\mu}^{0}, \varphi_{\tau}^{0}\}$, and the transfer matrix
$\mathcal{X}$ that maps $\Phi^{0}$ to $\Phi$ is not, in general,
invertible \footnote{We use the symbols $\Phi$ and $\varphi$ to denote the sum of neutrinos and antineutrinos except when we discuss implications of possible \textsf{CPT} violation, where $\Phi, \varphi$ refer to neutrinos, and $\bar{\Phi}, \bar{\varphi}$ to antineutrinos.}.  Moreover, over their long flight paths, cosmic neutrinos
are vulnerable to decay processes that would have gone undetected in
terrestrial or solar experiments.  Should the neutrino sector exhibit 
departures from \textsf{CPT} invariance, the pattern of
neutrino mixing could be considerably richer---and less circumscribed by
experiment---than conventional wisdom holds.

In this paper, we investigate the consequences of neutrino mixing for
the cosmic neutrino flux $\Phi$ at Earth, and explore how measurements
on Earth can characterize the source flux $\Phi^{0}$.  We contrast the
fluxes that
result from neutrino mixing  with those that might arise from simple decay scenarios, and we look at the unconventional consequences
that might obtain if \textsf{CPT} symmetry were violated.  We first carry out an
idealized analysis, taking our cue from current experiments; then we
take into account the uncertainties of existing experimental
constraints; finally we ask what we will know after the next round of
neutrino oscillation experiments. Our aim is to examine the scientific potential of neutrino observatories in light of current information about neutrino properties, and to project the situation five years hence.

The basic operational goal of neutrino telescopes is the detection of energetic---hence, long-range---muons generated in charged-current interactions $(\nu_{\mu},\bar{\nu}_{\mu})N \rightarrow (\mu^-,\mu^+)+\hbox{anything}$. Efficient, well-calibrated detection of $(\nu_e,\bar{\nu}_e)$ interactions is also required to make neutrino observatories incisive tools for the investigation of cosmic sources and neutrino properties. Good detection of $(\nu_{\tau},\bar{\nu}_{\tau})$ interactions would test the expectation that muon neutrinos and tau neutrinos arrive at Earth in nearly equal numbers. The normalized electron-neutrino flux at Earth emerges as a promising diagnostic for the character of cosmic sources and for nonstandard neutrino properties. If neutrinos behave as expected---mixing according to the standard three-generation picture---it should prove possible, over time, to infer something about the flavor mix of neutrinos produced in distant sources.
As is by now well known, neutrino oscillations---conventionally understood---map the standard neutrino mixture at the source, $\Phi^{0}_{\mathrm{std}} = \{\cfrac{1}{3},\cfrac{2}{3},0\}$ into a mixture at Earth that approximates $\{\cfrac{1}{3},\cfrac{1}{3},\cfrac{1}{3}\}$. We explore the uncertainties that attach to this expectation in light of recent improvements in the experimental constraints on neutrino-mixing parameters, and we compute the mixture at Earth to be expected for other source mixtures. Neutrino decays over the long path from astrophysical sources could distort the flavor mixture of neutrinos arriving at Earth. Current speculations about \textsf{CPT} violation in the neutrino sector imply small, and perhaps undetectable, modifications to the conventional oscillation scenario for ultrahigh-energy neutrinos, but would lead to striking consequences for antineutrino decay.

\section{The Influence of Neutrino Oscillations \label{oscil}}

Neutrino emission from active galactic nuclei (AGNs) may constitute
the dominant diffuse flux at energies above a few TeV, where cosmic
sources should emerge from the background of atmospheric neutrinos. 
With luminosities on the order of $10^{45\pm3}$~erg/s, AGNs are the
most powerful radiation sources in the universe.  They are cosmic
accelerators powered by the gravitational energy of matter falling in
upon a supermassive black hole.  Protons accelerated to very high
energies within an AGN may interact with ultraviolet photons in the
bright jets along the rotation axis or with matter in the accretion
disk.  The resulting $pp$ or $p\gamma$ collisions yield approximately equal numbers
of $\pi^+,\pi^0,\pi^-$ that decay into $\mu^{+}\nu_{\mu}$,
$\gamma\gamma$, or $\mu^{-}\bar{\nu}_{\mu}$.  The subsequent muon
decays yield additional muon neutrinos and electron neutrinos, so the
final count from $\pi^+ + \pi^0 + \pi^-$ is $2 \gamma + 2 \nu_{\mu} + 2
\bar{\nu}_{\mu} + 1 \nu_e + 1 \bar{\nu}_e$, for a normalized neutrino
flux 
\begin{equation}
\Phi^{0}_{\mathrm{std}} = \{\varphi_{e}^{0} = \cfrac{1}{3},
\varphi_{\mu}^{0} = \cfrac{2}{3}, \varphi_{\tau}^{0} = 0\}
\label{stdsource}
\end{equation}
 at the
source.  This standard mixture---equally divided between neutrinos 
and antineutrinos---is the canonical expectation from most
hypothesized sources of ultrahigh-energy neutrinos.

In the standard, three-generation description of neutrino oscillations, the flavor eigenstates $\nu_{\alpha}$ are related to the mass eigenstates $\nu_i$ through $\nu_{\alpha} = \sum_i U_{\beta i}\nu_i$, or
\begin{equation}
\pmatrix{\nu_e \cr \nu_{\mu} \cr \nu_{\tau}} = U \pmatrix{\nu_1 \cr \nu_2 \cr \nu_3} \; .
\end{equation}
A canonical form for  the neutrino mixing matrix is \cite{FKM}
\begin{equation}
    U = \pmatrix{U_{e1} & U_{e2} & U_{e3} \cr
U_{\mu1} & U_{\mu 2} & U_{\mu 3} \cr
U_{\tau1} & U_{\tau2} & U_{\tau3}} 
=
\pmatrix{c_{12}c_{13} & s_{12}c_{13} & s_{13}e^{-i\delta} \cr
-s_{12}c_{23} - c_{12}s_{23}s_{13}e^{i\delta} & c_{12}c_{23} - 
s_{12}s_{23}s_{13}e^{i\delta} & s_{23}c_{13} \cr
s_{12}s_{23} - c_{12}c_{23}s_{13}e^{i\delta} &  -c_{12}s_{23} - 
s_{12}c_{23}s_{13}e^{i\delta} & c_{23}c_{13}} \; ,
\end{equation}
where $s_{ij} = \sin\theta_{ij}$, $c_{ij} = \cos\theta_{ij}$ and $\delta$ is a \textsf{CP}-violating phase. A useful idealized form of the mixing matrix that holds for $\sin\theta_{13} = 0$, and maximal
atmospheric mixing, $\sin 2\theta_{23} = 1$, is
\begin{equation}
U_{\mathrm{ideal}} = \pmatrix{ c_{12} & s_{12} & 0 \cr
-s_{12}/\sqrt{2} & c_{12}/\sqrt{2} & 1/\sqrt{2} \cr
s_{12}/\sqrt{2} & -c_{12}/\sqrt{2} & 1/\sqrt{2}} \; .\label{Uideal}
\end{equation}

We can express the oscillation (or survival) probabilities in a transfer matrix $\mathcal{X}$ that maps an initial mixture of flavors at the source into the observed mixture at the detector,
\begin{equation}
\pmatrix{\varphi_e \cr \varphi_{\mu} \cr \varphi_{\tau}} = \mathcal{X} \pmatrix{\varphi_e^0 \cr \varphi_{\mu}^0 \cr \varphi_{\tau}^0}\; .
\end{equation}
Averaged over many oscillations, $\mathcal{X}$ takes the form 
\begin{eqnarray}
\mathcal{X}_{\beta\alpha} &=& \delta_{\alpha \beta} -
 2 \Re\sum_{i>j} U_{\alpha i}^*  U_{\beta i} U_{\alpha j} 
U_{\beta j}^* \nonumber \\
&=& \sum_{j} \mid U_{\alpha j}\mid^2 \;\; \mid
 U_{\beta j} \mid^2\; ,
\label{pro}
\end{eqnarray}
which does not depend on the phase $\delta$. The idealized form that corresponds to (\ref{Uideal}) is
\begin{equation}
\mathcal{X}_{\mathrm{ideal}} = \left(  
\begin{array}{ccc}
1-2x & x & x \\
x & \textstyle{\frac{1}{2}}(1-x) & \textstyle{\frac{1}{2}}(1-x) \\
x & \textstyle{\frac{1}{2}}(1-x) & \textstyle{\frac{1}{2}}(1-x)
\end{array}
\right)\; ,\label{idealtrans}
\end{equation}
where $x = \sin^{2}\theta_{12}\cos^{2}\theta_{12}$.
Because  the second and third rows are identical, the $\nu_{\mu}$ and $\nu_{\tau}$ fluxes that result from any source mixture $\Phi^{0}$ are equal: $\varphi_{\mu} = \varphi_{\tau}$. 

\begin{figure}[tb]
\vspace{6pt}
\centerline{\includegraphics[scale=0.45]{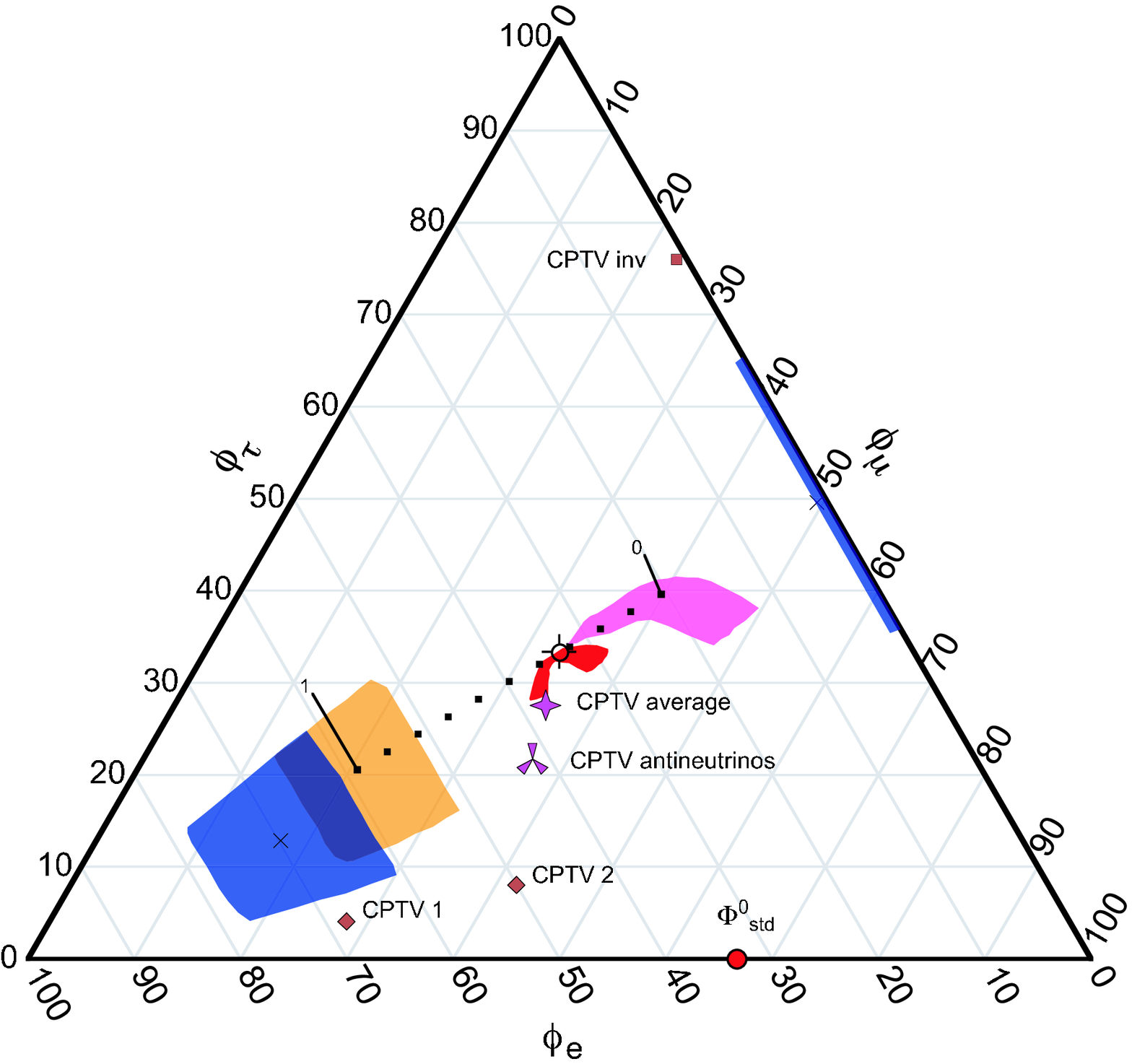}\quad\includegraphics[scale=0.45]{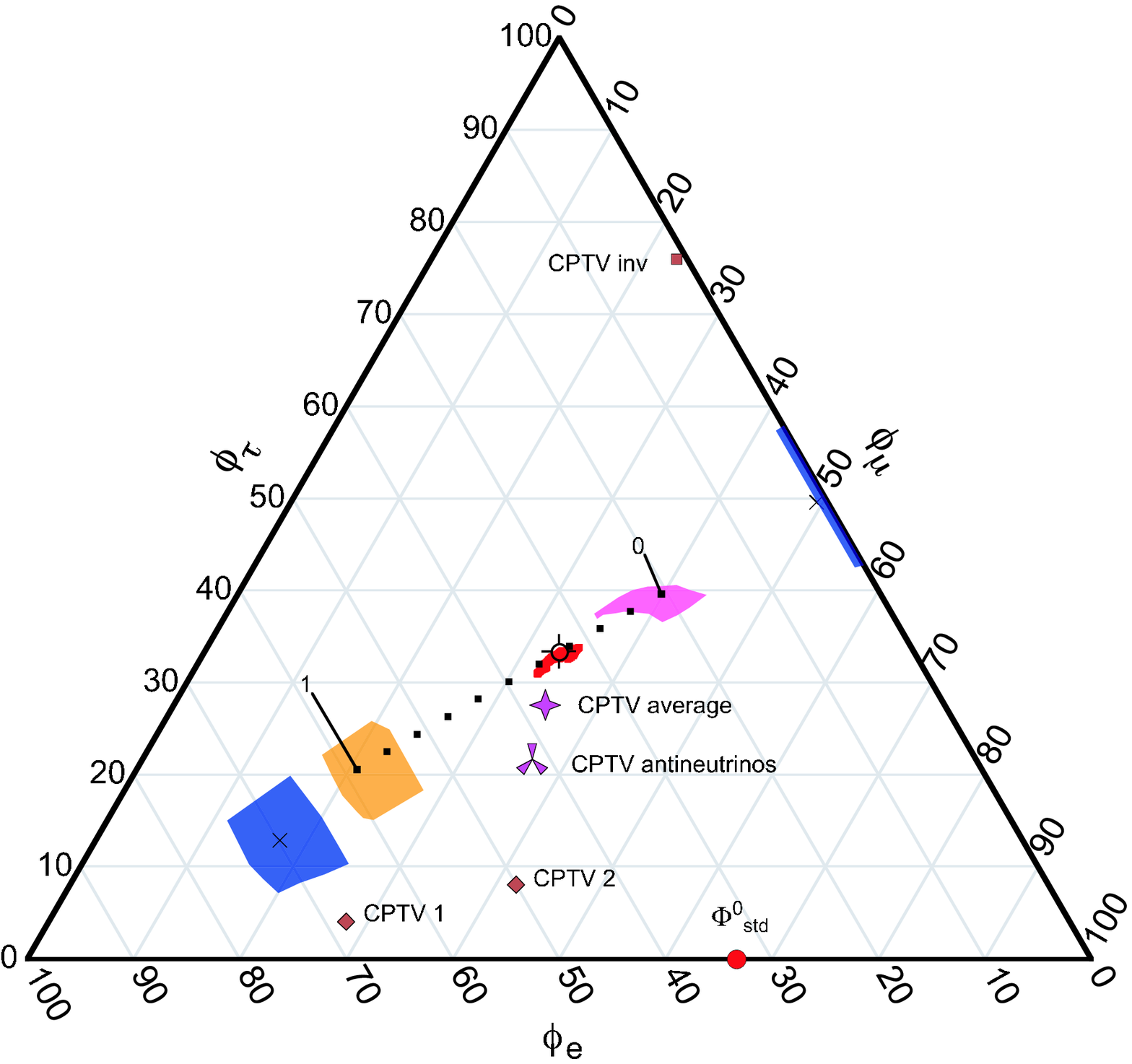}}
\caption{Ternary plots of the neutrino flux $\Phi$ at Earth, showing the implications of current (left pane) and future (right pane) knowledge of neutrino mixing.
The small black squares indicate 
the $\nu_{e}$ fractions produced by the idealized transfer matrix 
$\mathcal{X}_{\mathrm{ideal}}$ as $\varphi_{e}^{0}$ varies from 0 to 
1 in steps of 0.1. A crossed circle marks the standard mixed spectrum at Earth, 
$\Phi_{\mathrm{std}} = \{\cfrac{1}{3},\cfrac{1}{3},\cfrac{1}{3}\}$;
for comparison, a red dot marks the standard source spectrum,
$\Phi_{\mathrm{std}}^{0} = \{\cfrac{1}{3},\cfrac{2}{3},0\}$. Colored swaths delimit the fluxes at Earth produced by neutrino oscillations from the source mixtures $\Phi_0^0 = \{0,1,0\}$ (pink), $\Phi_{\mathrm{std}}^0$ (red), and $\Phi_1^0 = \{1,0,0\}$ (orange), using 95\% CL ranges for the oscillation parameters. Black crosses ($\times$) show the mixtures at Earth that follow from neutrino decay, assuming normal ($\varphi_e \approx 0.7$) and inverted ($\varphi_e \approx 0$) mass hierarchies. The blue bands show how current and future uncertainties blur the predictions for neutrino decays. The violet tripod indicates how \textsf{CPT}-violating oscillations shape the mix of antineutrinos that originate in a standard source mixture, and the violet cross averages that $\bar{\nu}$ mixture with the standard neutrino mixture. The brown squares denote consequences of \textsf{CPT} violation for antineutrino decays.
\label{nowdraft}}
\vspace{-6pt}
\end{figure}

It is noteworthy 
that $\mathcal{X}_{\mathrm{ideal}}$ maps the standard source mixture $\Phi^{0}_{\mathrm{std}}$ 
into 
\begin{equation}
    \Phi_{\mathrm{std}} = \{\varphi_{e} = \cfrac{1}{3},
\varphi_{\mu} = \cfrac{1}{3}, \varphi_{\tau} = \cfrac{1}{3}\},
\label{stdEarth}
\end{equation}
independent of $x$, i.e., independent of $\theta_{12}$.  Because $\nu_{\mu}$ and $\nu_{\tau}$ are fully
mixed, the result is more general: \textit{any source} with
$\varphi_{e}^{0} = \cfrac{1}{3}$ is mapped into $\Phi_{\mathrm{std}}$,
independent of $x$. The source mixture $\Phi_{Z}^{0} = \{\cfrac{1}{3},\cfrac{1}{3},\cfrac{1}{3}\}$ that arises from $Z^{0}$ decay is yet more special because it is unchanged by neutrino oscillations: \textit{any three-generation transfer matrix} maps it into $\Phi_{\mathrm{std}}$.
 For an arbitrary value of the $\nu_{e}$ fraction
at the source, $\mathcal{X}_{\mathrm{ideal}}$ leads
to $\varphi_{e} = \varphi_{e}^{0}(1 - 3x) + x$, with $\varphi_{\mu} =
\varphi_{\tau} = \cfrac{1}{2}(1 - \varphi_{e})$.  The variation of
$\varphi_{e}$ with the $\nu_{e}$ source fraction $\varphi_{e}^{0}$ is shown as a sequence of small black squares (for $\varphi_e^0 = 0, 0.1, \ldots , 1$) in Figure~\ref{nowdraft} for the value $x =
0.21$, which corresponds to $\theta_{12} = 0.57$, the central value in a recent global analysis~\cite{Alexei}.
The $\nu_{e}$ fraction at Earth ranges from $0.21$, for 
$\varphi_{e}^{0} = 0$, to $0.59$, for $\varphi_{e}^{0} = 1$. 

The simple analysis based on $\mathcal{X}_{\mathrm{ideal}}$ is useful for orientation, but it is important to explore the range of expectations implied by global fits to neutrino-mixing parameters. The recent KamLAND data \cite{KamLAND} on the reactor-$\bar{\nu}_e$ rate and spectral shape, taken together with solar-neutrino data, select the large-mixing-angle (LMA) solution, with a mixing angle bounded at 95\% CL to lie in the interval $0.49 <  \theta_{12} < 0.67$~\cite{Alexei}. Observations of atmospheric neutrinos favor maximal mixing; we take $\cfrac{\pi}{4}\times 0.8 < \theta_{23} < \cfrac{\pi}{4} \times 1.2$~\cite{Concha}, also at 95\% CL. Experimental evidence favors $\theta_{13} \ll 1$; informed by the nonobservation of oscillations in the CHOOZ and Palo Verde reactor experiments~\cite{Apollonio:2003gd,Boehm:2001ik}, we take  $0 < \theta_{13} < 0.1$. 

With current uncertainties in the oscillation parameters, a standard source spectrum, $\Phi_{\mathrm{std}}^{0} = \{\cfrac{1}{3},\cfrac{2}{3},0\}$, is mapped by oscillations onto the red boomerang near $\Phi_{\mathrm{std}} = \{\cfrac{1}{3},\cfrac{1}{3},\cfrac{1}{3}\}$ in the left pane of Figure~\ref{nowdraft}. Given that $\mathcal{X}_{\mathrm{ideal}}$ maps $\Phi_{\mathrm{std}}^{0} \rightarrow \Phi_{\mathrm{std}}$ for any value of $\theta_{12}$, it does not come as a great surprise that the target region is of limited extent, and this is a familiar result~\cite{OYasuda}. The variation of $\theta_{23}$ away from $\cfrac{\pi}{4}$ breaks the identity $\varphi_{\mu}\equiv\varphi_{\tau}$ of the idealized analysis.

One goal of neutrino observatories will be to characterize cosmic sources by determining the source mix of neutrino flavors. It is therefore of interest to examine the outcome of different assumptions about the source. We show in the left pane of Figure~\ref{nowdraft} the mixtures at Earth implied by current knowledge of the oscillation parameters for source fluxes $\Phi_0^0 = \{0,1,0\}$ (the purple band near $\varphi_e \approx 0.2$) and $\Phi_1^0 = \{1,0,0\}$ (the orange band near $\varphi_e \approx 0.6$) \footnote{Muon cooling within the source, as discussed in Ref.~\cite{RachMesz}, would deplete the source flux of $\nu_e$. We have no candidate mechanism to produce an ultrahigh-energy flux like $\Phi_1^0$. In view of the current puzzlement over the origins of ultrahigh-energy cosmic rays, we think it prudent to keep an open mind about the range of possibilities.}. For the $\Phi^0_{\mathrm{std}}$ and $\Phi^0_1$ source spectra, the uncertainty in $\theta_{12}$ is reflected mainly in the variation of $\varphi_e$, whereas the uncertainty in $\theta_{23}$ is expressed in the variation of $\varphi_{\mu}/\varphi_{\tau}$ For the $\Phi_0^0$ case, the influence of the two angles is not so orthogonal. For all the source spectra we consider, the uncertainty in $\theta_{13}$ has little effect on the flux at Earth. The extent of the three regions, and the absence of a clean separation between the regions reached from $\Phi_{\mathrm{std}}^{0}$ and $\Phi_0^0$ indicates that characterizing the source flux will be challenging, in view of the current uncertainties of the oscillation parameters.

Over the next five years---roughly the time scale on which large-volume neutrino telescopes will come into operation---we can anticipate improved information on $\theta_{12}$ and $\theta_{23}$ from KamLAND \cite{Barger:2000hy} and the long-baseline accelerator experiments at Soudan \cite{MINOS} and Gran Sasso \cite{CNGS}. We base our projections for the future on the ranges $0.54 < \theta_{12} < 0.63$ and $\cfrac{\pi}{4} \times 0.9 < \theta_{23} < \cfrac{\pi}{4} \times 1.1$, still with $0 < \theta_{13} < 0.1$ The results are shown in the right panel of Figure~\ref{nowdraft}. The (purple) target region for the source flux $\Phi_0^0$ shrinks appreciably and separates from the (red) region populated by $\Phi_{\mathrm{std}}^0$, which is now tightly confined around $\Phi_{\mathrm{std}}$. The (orange) region mapped from the source flux $\Phi_1^0$ by oscillations shrinks by about a factor of two in the $\varphi_e$ and $\varphi_{\mu}-\varphi_{\tau}$ dimensions.

If \textsf{CPT} invariance is not respected in the neutrino sector~\cite{BBLideas}, the neutrino and antineutrino mixing matrices need not coincide. The KamLAND experiment's confirmation of solar-like oscillations of $\bar{\nu}_e$ restricts the space of \textsf{CPT}-violating mixing matrices, but Barenboim, Borissov, and Lykken~\cite{BBLKamLAND} have found two illustrative solutions that respect all existing constraints:
\begin{equation}
\hbox{Solution 1:} \pmatrix{\theta_{12} = 0.6  \cr \theta_{23}  = 0.5 \cr \theta_{13} = 0.08}\; ; \quad
\hbox{Solution 2:} \pmatrix{\theta_{12} = 0.785  \cr \theta_{23}  = 0.52 \cr \theta_{13} = 0.08}\; .
\label{bblsolns}
\end{equation}
These two solutions lead to essentially identical transfer matrices (cf.\ Eqn.\ (\ref{pro})) that map a standard source mixture of \textit{anti}neutrinos into $\bar{\Phi}_{\mathsf{CPTV}} = \{0.42, 0.36, 0.22\}$, which is plotted as a violet tripod in Figure~\ref{nowdraft}. Observing $\varphi_e/\bar{\varphi}_e = \cfrac{4}{5} \neq 1$ would be manifest evidence of \textsf{CPT} violation. Although in principle it might be possible to distinguish the flux of $\nu_e$ and $\bar{\nu}_e$ at Earth by observing the resonant formation $\bar{\nu}_e e \rightarrow W^- \rightarrow \hbox{hadrons}$, the $\nu_e + \bar{\nu}_e$ flux is more likely to be the available observable. In the presence of \textsf{CPT}-violating mixing, the expectation from a standard flux at the source is $\langle \Phi_{\mathsf{CPTV}} \rangle = \cfrac{1}{2} (\Phi_{\mathrm{std}} + \bar{\Phi}_{\mathsf{CPTV}} ) = \{0.375, 0.35, 0.275\}$,
which is plotted as a violet cross in Figure~\ref{nowdraft}. Distinguishing this value from $\Phi_{\mathrm{std}} = \{\cfrac{1}{3}, \cfrac{1}{3}, \cfrac{1}{3}\}$ would be extraordinarily challenging.

\section{Reconstructing the Neutrino Mixture at the Source}
\begin{figure}[t!]
\vspace{6pt}
\centerline{\includegraphics[scale=0.5]{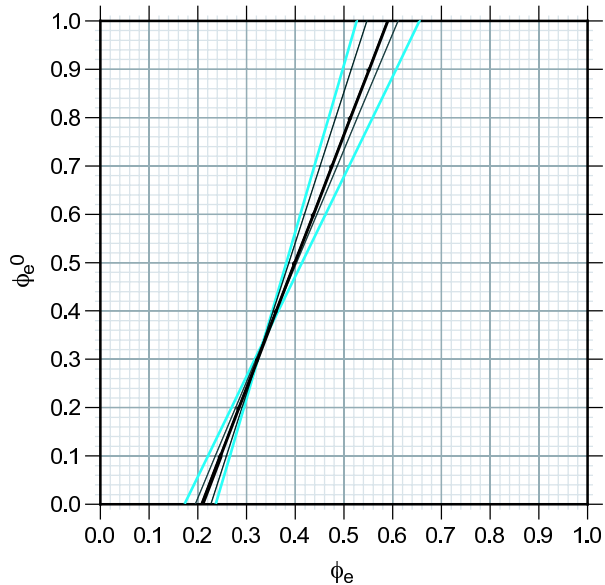}}
\caption{The source flux $\varphi_e^0$ of electron neutrinos reconstructed from the $\nu_e$ flux $\varphi_e$ at Earth, using the ideal transfer matrix $\mathcal{X}_{\mathrm{ideal}}$ of Eqn.\ (\ref{idealtrans}). The heavy solid line refers to our chosen central value, $\theta_{12} = 0.57$. The light blue lines refer to the current experimental constraints (at 95\% CL), and the thin solid lines refer to our projection of future experimental constraints.
\label{inversion}}
\vspace{-6pt}
\end{figure}

What can observations of the blend $\Phi$ of neutrinos arriving at
Earth tell us about the source?  Inferring the nature of the processes that generate cosmic neutrinos is more complicated than it would be in the absence of neutrino oscillations. Because $\nu_{\mu}$ and $\nu_{\tau}$
are fully mixed---and thus enter identically in
$\mathcal{X}_{\mathrm{ideal}}$---it is not possible fully to characterize
$\Phi^{0}$. We can, however, reconstruct the $\nu_{e}$ fraction at the 
source as
\begin{equation}
    \varphi_{e}^{0} = \frac{\varphi_{e} - x}{1 - 3x}\; .
\end{equation}
The reconstructed source flux $\varphi_e^0$ is shown in Figure~\ref{inversion} as a function of the $\nu_e$ flux at Earth. The heavy solid line represents the best-fit value for $\theta_{12}$; the light blue lines and thin solid lines indicate the current and future 95\% CL bounds on $\theta_{12}$.

A possible strategy for beginning to characterize a source of cosmic neutrinos might proceed by measuring the $\nu_e/\nu_{\mu}$ ratio and estimating $\varphi_e$ under the plausible assumption---later to be checked---that $\varphi_{\mu} = \varphi_{\tau}$. Let us first note that very large ($\varphi_e \agt 0.65$) or very small ($\varphi_e \alt 0.15$) $\nu_e$ fluxes cannot be accommodated in the standard neutrino-oscillation picture. Observation of an extreme $\nu_e$ fraction would implicate unconventional physics.

As we have already seen from Figure~\ref{nowdraft}, constraining the source flux sufficiently to test the nature of the neutrino production process will require rather precise determinations of the neutrino flux at Earth. Suppose we want to test the idea that the source flux has the standard composition of Eqn.~(\ref{stdsource}).  With today's uncertainty on $\theta_{12}$, a 30\% measurement that locates $\varphi_e = 0.33 \pm 0.10$ implies only that $0 \alt \varphi_e^0 \alt 0.68$. For a measured flux in the neighborhood of $\cfrac{1}{3}$, the uncertainty in the solar mixing angle is of little consequence: the constraint that arises if we assume the central value of $\theta_{12}$ is not markedly better: $0.06 \alt \varphi_e^0 \alt 0.59$. A 10\% measurement of the $\nu_e$ fraction, $\varphi_e = 0.33 \pm 0.033$, would make possible a rather restrictive constraint on the nature of the source. The central value for $\theta_{12}$ leads to $0.26 \alt \varphi_e^0 \alt 0.43$, blurred to $0.22 \alt \varphi_e^0 \alt 0.45$ with current uncertainties.

Measured $\nu_e$ fractions that depart significantly from the canonical $\varphi_e = \cfrac{1}{3}$ would suggest nonstandard neutrino sources. An observed flux $\varphi_e = 0.5 \pm 0.1$ points to a source flux $0.47 \alt \varphi_e^0 \alt 1$, with current uncertainties, whereas $\varphi_e = 0.25 \pm 0.10$ indicates $0 \alt \varphi_e^0 \alt 0.35$.

\section{Influence of Neutrino Decays}
Beacom, Bell, and collaborators \cite{BBHPW,Beacom:2002cb} have observed that the decays of unstable neutrinos over cosmic distances can lead to mixtures at Earth that are incompatible with the oscillations of stable neutrinos. The candidate decays are transitions between mass eigenstates by emission of a very light particle, $\nu_i \rightarrow (\nu_j, \bar{\nu}_j)+X$. Dramatic effects occur when the decaying neutrinos disappear, either by decay to invisible products or by decay into active neutrino species so degraded in energy that they contribute negligibly to the total flux at the lower energy. 
If the lifetimes of the unstable mass eigenstates are short compared with the flight time from source to Earth, decay of the unstable neutrinos will be complete, and the (unnormalized) flavor $\nu_{\alpha}$ flux at Earth is given by
\begin{equation}
\widetilde{\varphi}_{\alpha} (E_{\nu})=  
\sum_{i = \mathrm{stable} } \sum_{\beta} \varphi^{0}_{\beta}(E_{\nu})
|U_{\beta i}|^2 |U_{\alpha i}|^2\;\;,
\label{fulldecay}
\end{equation}
with $\varphi_{\alpha} = \widetilde{\varphi}_{\alpha} / \sum_{\beta} \widetilde{\varphi}_{\beta}$.

If only the lightest neutrino survives, the flavor mix of neutrinos arriving at Earth is determined by the flavor composition of the lightest mass eigenstate, \textit{independent of the flavor mix at the source.} (The flavor mix at the source of course determines the relative fluxes of the mass eigenstates at the source, and so the rate of surviving neutrinos at Earth.)
For a normal mass hierarchy $m_1 < m_2 < m_3$,  the $\nu_{\alpha}$ flux at Earth is $\varphi_{\alpha} =  |U_{\alpha 1}|^2$. Accordingly, the neutrino flux at Earth is $\Phi_{\mathrm{normal}} = \{|U_{e1}|^2, |U_{\mu1}|^2,|U_{\tau1}|^2\} \approx  \{0.70, 0.17, 0.13\}$ for our chosen central values of the mixing angles. If the mass hierarchy is inverted, $m_1 > m_2 > m_3$, the lightest (hence, stable) neutrino is $\nu_3$, so the flavor mix at Earth is determined by $\varphi_{\alpha} = |U_{\alpha 3}|^2$. In this case, the neutrino flux at Earth is $\Phi_{\mathrm{inverted}} = \{|U_{e3}|^2, |U_{\mu3}|^2,|U_{\tau3}|^2\} \approx \{0, 0.5, 0.5\}$. Both $\Phi_{\mathrm{normal}}$ and $\Phi_{\mathrm{inverted}}$, which are indicated by crosses ($\times$) in Figure~\ref{nowdraft}, are very different from the standard flux $\Phi_{\mathrm{std}} = \{\varphi_{e} = \cfrac{1}{3}, \varphi_{\mu} = \cfrac{1}{3}, \varphi_{\tau} = \cfrac{1}{3}\}$ produced by the ideal transfer matrix from a standard source. Observing either mixture would represent a departure from conventional expectations.

The fluxes that result from neutrino decays \textit{en route} from the sources to Earth are subject to uncertainties in the neutrino-mixing matrix. The expectations for the two decay scenarios are indicated by the blue regions in Figure~\ref{nowdraft}, where we indicate the consequences of 95\% CL ranges of the mixing parameters now and in the future. With current uncertainties, the normal hierarchy populates $0.61 \alt \varphi_e \alt 0.77$, and allows considerable departures from $\varphi_{\mu} = \varphi_{\tau}$. The normal-hierarchy decay region based on current knowledge overlaps the flavor mixtures that oscillations produce in a pure-$\nu_e$ source, shown in orange. (It is, however, far removed from the standard region that encompasses $\Phi_{\mathrm{std}}$.) With the projected tighter constraints on the mixing angles, the range in $\varphi_e$ swept out by oscillation from a pure-$\nu_e$ source or decay from a normal hierarchy shrinks by about a factor of two. Neutrino decay then populates $0.65 \alt \varphi_e \alt 0.74$, and is separated from the oscillations. The degree of separation between the region populated by normal-hierarchy decay and the one populated by mixing from a pure-$\nu_e$ source depends on the value of the solar mixing angle $\theta_{12}$. For the seemingly unlikely value $\theta_{12} = \cfrac{\pi}{4}$, both mechanisms yield $\Phi = \{\cfrac{1}{2}, \cfrac{1}{4}, \cfrac{1}{4}\}$.

With current uncertainties, the inverted hierarchy spans the range $(\varphi_{\mu}, \varphi_{\tau}) = (0.34, 0.66)\hbox{ to }(0.66, 0.34)$, always with $\varphi_e\approx 0$. With the improvements we project in the knowledge of mixing angles, the range will decrease to $(\varphi_{\mu}, \varphi_{\tau}) = (0.42, 0.58)\hbox{ to }(0.58, 0.42)$, with $\varphi_e\approx 0$. In both cases, these mixtures are well separated from the mixtures that would result from neutrino oscillations, for any conceivable source at cosmic distances.

Should \textsf{CPT} not be a good symmetry for neutrinos, the mass hierarchies and mixing matrices can be different for neutrinos and antineutrinos. Different numbers of stable $\nu$ and $\bar{\nu}$ may reach the Earth, so we do not know how to combine neutrino and antineutrino results to compare with neutrino-telescope measurements. The analysis of neutrino decays does not change if we relax \textsf{CPT} invariance, because neutrino mixing is already highly constrained by experiment. The Barenboim--Borissov--Lykken antineutrino mixing parameters  \cite{BBLKamLAND} (cf.\ Eqn.\ (\ref{bblsolns}))  provide examples of what can be expected for antineutrino decays if \textsf{CPT} is violated. Consider first the normal mass hierarchy, in which $\nu_2$ and $\nu_3$ both decay completely before reaching Earth. The resulting antineutrino fluxes at Earth are $\bar{\Phi}_{\mathsf{CPTV1}} = \{0.68, 0.28, 0.04\}$ and $\bar{\Phi}_{\mathsf{CPTV2}} = \{0.50, 0.42, 0.08\}$, both characterized by unusually large ratios of $\bar{\varphi}_{\mu}/\bar{\varphi}_{\tau}$ that place them outside the range of \textsf{CPT}-conserving decays and oscillations alike \footnote{The barred quantities refer to antineutrinos.}. In the case of an inverted mass hierarchy, both BBL parameter sets predict $\bar{\Phi}_{\mathsf{CPTV~inv}} = \{0.01, 0.23, 0.76\}$, an unusually small ratio of $\bar{\varphi}_{\mu}/\bar{\varphi}_{\tau}$. Conventional oscillations---or \textsf{CPT}-conserving decays---populate the ternary plots of Figure \ref{nowdraft} along the line $\varphi_{\mu} = \varphi_{\tau}$. \textsf{CPT}-violating decays offer an example of exotic physics with a very specific signature.

Beacom and Bell~\cite{Beacom:2002cb} have shown that observations of solar neutrinos set the most stringent plausible lower bound on the reduced lifetime of a neutrino of mass $m$ as $\tau/m \agt 10^{-4}\hbox{ s/eV}$. This rather modest limit opens the possibility that some neutrinos do not survive the journey from astrophysical sources, with consequences we have just explored. The energies of neutrinos that may be detected in the future from AGNs and other cosmic sources range over several orders of magnitude, whereas the distances to such sources vary over perhaps one order of magnitude. The neutrino energy sets the neutrino lifetime in the laboratory frame; more energetic neutrinos survive over longer flight paths than their lower-energy companions \footnote{A similar phenomenon is familiar for cosmic-ray muons.}. Under propitious circumstances of reduced lifetime, path length, and neutrino energy, it might be possible to observe the transition from more energetic survivor neutrinos to less energetic decayed neutrinos.

\begin{figure}[tb]
\vspace{6pt}
\centerline{\includegraphics[scale=0.5]{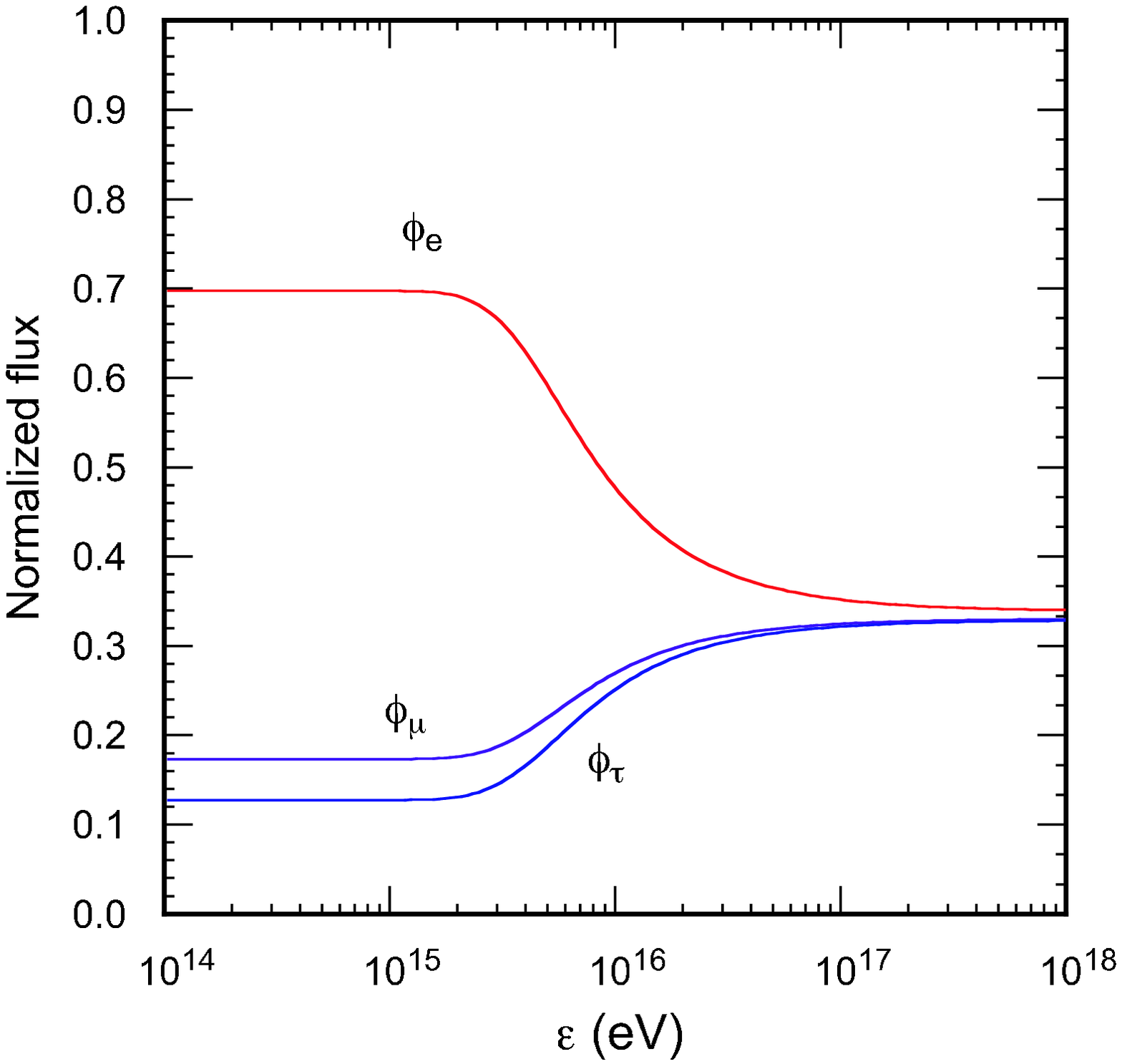}\quad\includegraphics[scale=0.5]{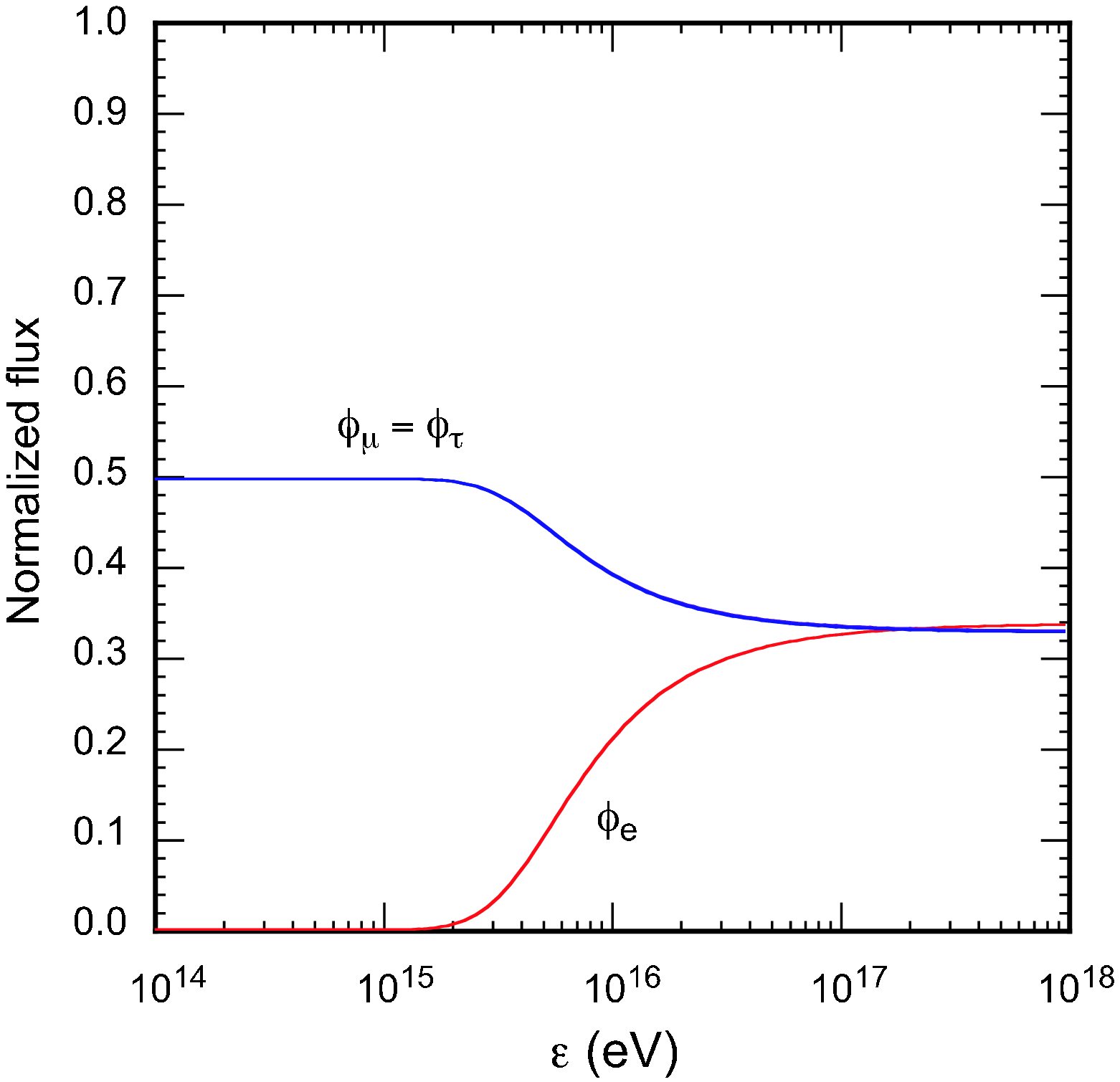}}
\caption{Energy dependence of normalized $\nu_e$, $\nu_{\mu}$, and $\nu_{\tau}$ fluxes, for the two-body decay of the two upper mass eigenstates, with the neutrino source at $L=100\hbox{ Mpc}$ from Earth and $\tau/m = 1\hbox{ s/eV}$. The left pane shows the result for a normal mass hierarchy; the right pane shows the result for an inverted mass hierarchy. With suitable rescaling of the neutrino energy (cf. Eqn. (\ref{rescale})), these plots apply for any combination of path length and reduced lifetime.
\label{propitious}}
\vspace{-6pt}
\end{figure}

If decay is not complete, the (unnormalized) flavor $\nu_{\alpha}$ flux arriving at Earth from a source at distance $L$ is given by
\begin{equation}
\widetilde{\varphi}_{\alpha} (E_{\nu})=  
\sum_{i } \sum_{\beta} \varphi^{0}_{\beta}(E_{\nu})
|U_{\beta i}|^2 |U_{\alpha i}|^2 e^{-(L/E_{\nu})(m_i/\tau_i)}\;\;,
\label{partialdecay}
\end{equation}
with the normalized flux $\varphi_{\alpha}(E_{\nu}) = \widetilde{\varphi}_{\alpha}(E_{\nu}) / \sum_{\beta} \widetilde{\varphi}_{\beta}(E_{\nu})$. An idealized case will illustrate the possibilities for observing the onset of neutrino decay and estimating the reduced lifetime. Assume a normal mass hierarchy, $m_1 < m_2 < m_3$, and let $\tau_3/m_3 = \tau_2/m_2 \equiv \tau/m$. For a given path length $L$, the neutrino energy at which the transition occurs from negligible decays to complete decays is determined by $\tau/m$. We show in the left pane of Figure~\ref{propitious} the energy evolution of the normalized neutrino fluxes arriving from a standard source; the energy scale is appropriate for the case $\tau/m = 1\hbox{ s/eV}$ and $L = 100\hbox{ Mpc}$. The result shown there is actually universal---within our simplifying assumptions---provided that the neutrino energy scale is renormalized as
\begin{equation}
E_{\nu} = \varepsilon \left( \displaystyle{\frac{1\hbox{ s/eV}}{\tau/m}}\right) \left(\frac{L}{100\hbox{ Mpc}}\right) . \label{rescale}
\end{equation}

\begin{figure}[b!]
\vspace{6pt}
\centerline{\includegraphics[scale=0.45]{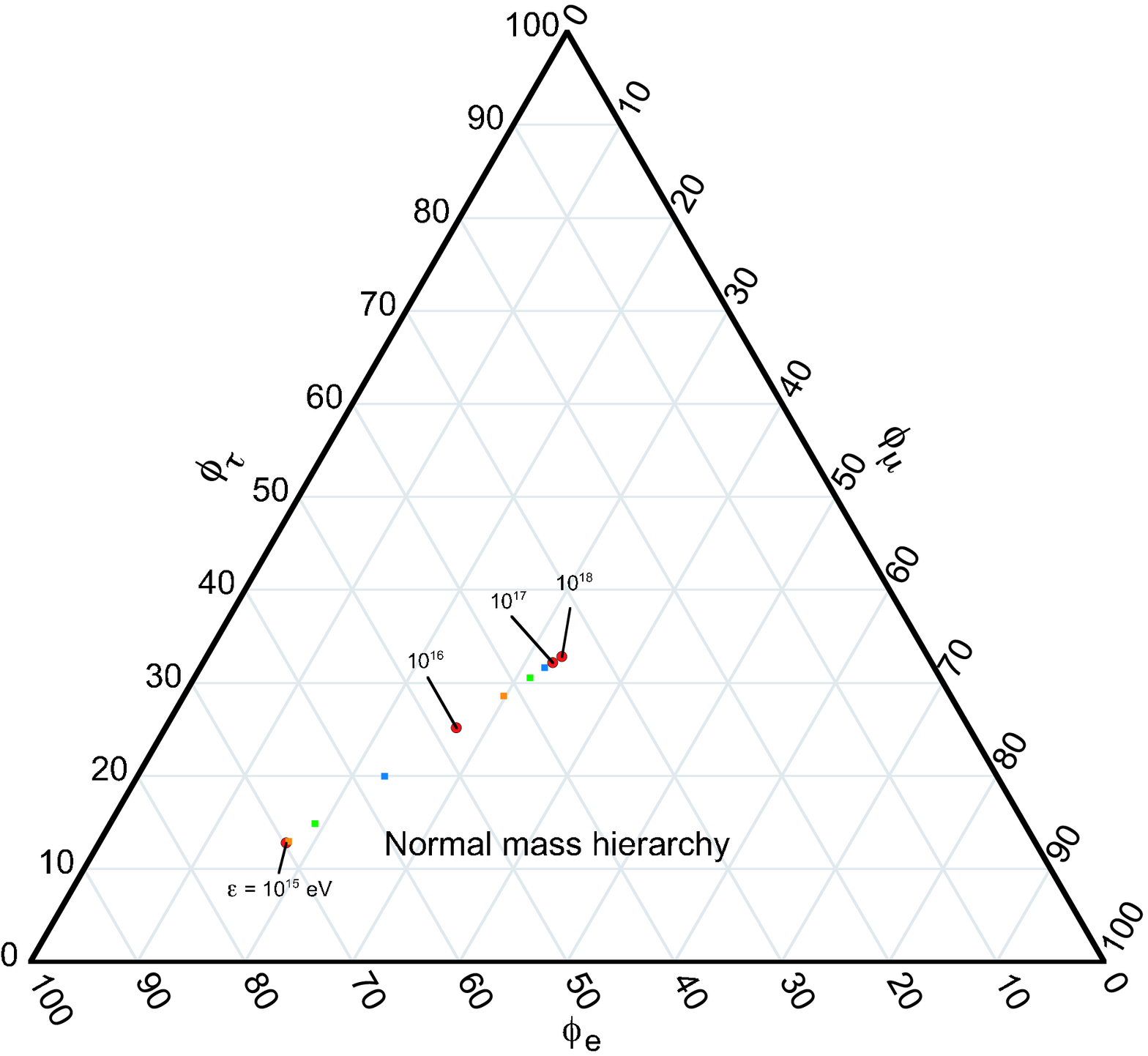}\quad\includegraphics[scale=0.45]{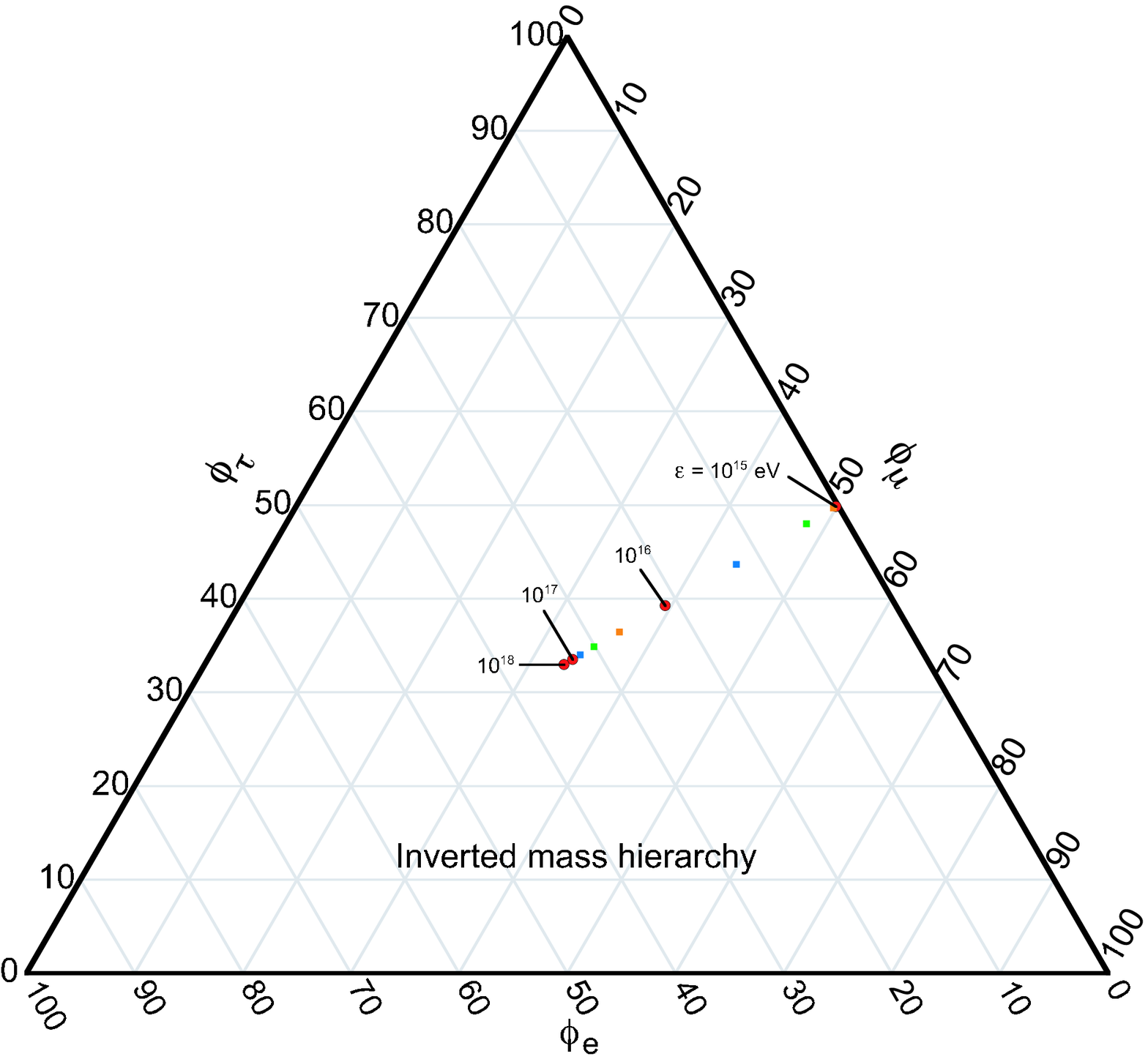}}
\caption{Ternary plots showing the energy dependence of normalized $\nu_e$, $\nu_{\mu}$, and $\nu_{\tau}$ fluxes, for the two-body decay of the two upper mass eigenstates, with the neutrino source at $L=100\hbox{ Mpc}$ from Earth and $\tau/m = 1\hbox{ s/eV}$. The left pane shows the result for a normal mass hierarchy; the right pane shows the result for an inverted mass hierarchy. Each unlabeled step multiplies $\varepsilon$ by $10^{1/4}$. With suitable rescaling of the neutrino energy (cf. Eqn. (\ref{rescale})), these plots apply for any combination of path length and reduced lifetime.
\label{propitious2}}
\vspace{-6pt}
\end{figure}

We show in  the left pane of Figure~\ref{propitious2} that over about one decade in energy, the flavor mix of neutrinos arriving at Earth changes from the normal decay flux, $\Phi_{\mathrm{normal}}$, to the standard oscillation flux, $\Phi_{\mathrm{std}}$. For the case at hand, the transition takes place between $3\times 10^{15}\hbox{ eV}$ and $3\times 10^{16}\hbox{ eV}$. The corresponding results for the inverted mass hierarchy are depicted in the right panes of Figures~\ref{propitious} and \ref{propitious2}, which show a rapid transition from $\Phi_{\mathrm{inverted}}$ to $\Phi_{\mathrm{std}}$. 

If we locate the transition from survivors to decays at neutrino energy $E^{\star}$, then we can estimate the reduced lifetime in terms of the distance to the source as
\begin{equation}
\tau / m \approx 100\hbox{ s/eV} \cdot \left(\frac{L}{\hbox{1 Mpc}}\right) \left( \frac{1\hbox{ TeV}}{E^{\star}}\right)\; .
\label{taumest}
\end{equation}
In practice, ultrahigh-energy neutrinos are likely to arrive from a multitude of sources at different distances from Earth, so the transition region will be blurred \footnote{Our assumption that $\tau_3/m_3 = \tau_2/m_2 \equiv \tau/m$ is also a special case.}. Nevertheless, it would be rewarding to observe the decay-to-survival transition, and to use Eqn.\ (\ref{taumest}) to estimate---even within one or two orders of magnitude---the reduced lifetime. If no evidence appears for a flavor mix characteristic of neutrino decay, then Eqn.\ (\ref{taumest}) provides a lower bound on the neutrino lifetime. For that purpose, the advantage falls to large values of $L/E^{\star}$, and so to the lowest energies at which neutrinos from distant sources can be observed. Observing the standard flux, $\Phi_{\mathrm{std}} = \{\cfrac{1}{3},\cfrac{1}{3},\cfrac{1}{3}\}$, which is incompatible with neutrino decay, would strengthen the current bound on $\tau/m$ by some seven orders of magnitude,  for 10-TeV neutrinos from sources at 100~Mpc~\cite{Beacom:2002cb}.

\section{Assessment}
The conventional mechanism for ultrahigh-energy neutrino production in astrophysical sources, when modulated by the standard picture of neutrino oscillations, leads to approximately equal fluxes of $\nu_e$, $\nu_{\mu}$, and $\nu_{\tau}$ at a terrestrial detector. Observing a strong deviation from this expectation will indicate that we must revise our conception of neutrino sources---or that neutrinos behave in unexpected ways on the long journey from source to Earth. With the more precise understanding of neutrino mixing that we expect will develop over the next five years, it may become possible to characterize the neutrino mixture at the source. On the other hand, detecting equal fluxes of the three neutrino flavors can strengthen existing bounds on neutrino lifetimes quite dramatically. If neutrinos do decay between source and Earth, the flavor mixture at Earth will be incompatible with the consequences of the standard source mixture. As constraints improve on the mixing angles $\theta_{12}$ and $\theta_{23}$, the flavor mix that results from neutrino decay will separate decisively from what can be generated by any source mix plus oscillations. In the best imaginable case for neutrino decays, the neutrino lifetime might reveal itself in an energy dependence of the flavor mix observed at Earth. If neutrinos decay, the fluxes at Earth are very different for normal and inverted mass hierarchies.

Conventional neutrino oscillations---and conventional decays---lead to nearly equal fluxes of $\nu_{\mu}$ and $\nu_{\tau}$, and so populate only a portion of the $\varphi_e$--$\varphi_{\mu}$--$\varphi_{\tau}$ ternary plot. A neutrino mixture at Earth that is far from the $\varphi_{\mu}=\varphi_{\tau}$ line points to unconventional neutrino physics. \textsf{CPT} violation offers one example of novel behavior, but its predictions are tested most cleanly by observing differences between neutrinos and antineutrinos, for which the experimental prospects are distinctly limited at ultrahigh energies.

The analysis presented here offers further motivation to develop techniques for identifying interactions of ultrahigh-energy $\nu_e$, $\nu_{\mu}$, and $\nu_{\tau}$---and for measuring the neutrino energies---in cubic-kilometer--scale neutrino observatories. If this can be achieved, prospects for learning about the properties of neutrinos and about the nature of cosmic neutrino sources will be greatly enhanced.
\begin{acknowledgments}
We thank Stephen Parke for a helpful observation, and thank John Beacom and Nicole Bell for lively discussions.
Fermilab is operated by Universities Research Association Inc.\ under
Contract No.\ DE-AC02-76CH03000 with the U.S.\ Department of Energy. One of us (C.Q.) is grateful for the hospitality of the Kavli Institute for Theoretical Physics during the program, \textit{Neutrinos:Ê Data, Cosmos, and Planck Scale.} This research was supported in part by the National Science Foundation under Grant No.\ PHY99-07949.

\end{acknowledgments}


\begin{thebibliography}{99}

    \bibitem{MannLear} J. G. Learned and K. Mannheim, \textit{Annu.  Rev. 
    Nucl.  Part.  Sci.} \textbf{50,} 679-749 (2000).
    
    \bibitem{francis} F. Halzen and D. Hooper, \textit{Rept.  Prog. 
    Phys.} \textbf{65,} 1025 (2002).

    \bibitem{francis03} F. Halzen, ``High-Energy Neutrino Astronomy: Science and First Results,'' \textsf{[arXiv:astro-ph/0301143]}.
    
    \bibitem{GQRSapp} R. Gandhi, C. Quigg, M. H. Reno, and I. Sarcevic,
    \textit{Astropart.  Phys.} \textbf{5,} 81-110 (1996); 
    \textit{Phys.  Rev.  D}\textbf{58,} 093009 (1998).
    
    \bibitem{FKM} Peter Fisher, Boris Kayser, and Kevin S. McFarland, \textit{Annu. Rev. Nucl. Part. Sci.} \textbf{49,} 481-527 (1999).

\bibitem{Alexei} P. C. de Holanda and A. Yu. Smirnov,  ``LMA MSW Solution of the Solar Neutrino Problem and First KamLAND Results,'' \textsf{[arXiv:hep-ph/0212270]}.

\bibitem{KamLAND} K. Eguchi, et al.\ (KamLAND Collaboration), ``First Results from KamLAND:  Evidence for Reactor Antineutrino Disappearance,'' \textsf{[arXiv:hep-ex/0212021]}.

\bibitem{Concha}  M. C. Gonzalez-Garcia, ``Theory of Neutrino Masses and Mixing,'' 
Plenary report at the 2003 International Conference on High Energy Physics, Amsterdam, \textsf{[arXiv:hep-ph/0210359]}.

\bibitem{Apollonio:2003gd}
M.~Apollonio, {\it et al.} (CHOOZ Collaboration),
\textsf{[arXiv:hep-ex/0301017]}.

\bibitem{Boehm:2001ik}
F.~Boehm, {\it et al.},
\textit{Phys.\ Rev.\ D}{\bf 64}, 112001 (2001)
\textsf{[arXiv:hep-ex/0107009]}.

\bibitem{OYasuda} J. G. Learned and S. Pakvasa, \textit{Astropart. 
Phys.} \textbf{3,} 267 (1995); H. Athar, M. Jezabek, and O. Yasuda,
\textit{Phys.  Rev.  D}\textbf{62,} 103007 (2000).  Osamu Yasuda, ``Neutrino oscillations in high energy cosmic neutrino flux,''
\textsf{[arXiv:hep-ph/0005135]}; G. J. Gounaris and G. Moultaka, ``The Flavor Distribution of Cosmic Neutrinos,'' \textsf{[arXiv:hep-ph/0212110]}.

\bibitem{RachMesz} J. P. Rachen and P. M\'{e}sz\'{a}ros, \textit{Phys. 
Rev. D}\textbf{58,} 123005 (1998).


\bibitem{Barger:2000hy} V.~D.~Barger, D.~Marfatia, and B.~P.~Wood,
\textit{Phys.\ Lett.} {\bf B498}, 53 (2001);
see also  \url{http://www.awa.tohoku.ac.jp/KamLAND/}.

\bibitem{MINOS} \url{http://www-numi.fnal.gov}.

\bibitem{CNGS} \url{http://proj-cngs.web.cern.ch/proj-cngs/}.

\bibitem{BBLideas}
G.~Barenboim, L.~Borissov, J.~Lykken, and A.~Yu.~Smirnov,
\textit{JHEP} {\bf 0210}, 001 (2002)
\textsf{[arXiv:hep-ph/0108199]};
G.~Barenboim, L.~Borissov, and J.~Lykken,
\textit{Phys.\ Lett.\ } {\bf B534}, 106 (2002)
\textsf{[arXiv:hep-ph/0201080]}.
    
    \bibitem{BBLKamLAND}  G. Barenboim, L. Borissov, and J. Lykken,
``CPT Violating Neutrinos in the Light of KamLAND,'' \textsf{[arXiv:hep-ph/0212116]}.
 
    \bibitem{BBHPW}  J. F. Beacom, N. F. Bell, D. Hooper, S. Pakvasa, and 
    T. J. Weiler, ``Decay of High-Energy Astrophysical Neutrinos,'' \textsf{[arXiv:hep-ph/0211305]}.

\bibitem{Beacom:2002cb}
J.~F.~Beacom and N.~F.~Bell,
\textit{Phys.\ Rev.\ D}{\bf 65}, 113009 (2002)
\textsf{[arXiv:hep-ph/0204111]}.


\end{thebibliography}


\end{document}